\documentclass[aps,prl,twocolumn,showpacs,superscriptaddress]{revtex4-2}
\usepackage{graphicx}
\usepackage{amsmath}
\usepackage{amssymb}
\usepackage{colordvi}
\usepackage{mathrsfs}
\usepackage{bm}
\usepackage{verbatim}
\usepackage{dcolumn}
\usepackage{bm}
\usepackage{epsfig}
\usepackage{subfigure}
\usepackage{dsfont}
\usepackage{multirow}
\usepackage[colorlinks,linkcolor=blue,anchorcolor=blue,citecolor=blue]{hyperref}

\begin{document}
\title{Magnon-Driven Anomalous Hall Effect in Altermagnets}
\author{Zheng Liu}
\affiliation{CAS Key Laboratory of Strongly-Coupled Quantum Matter Physics, and Department of Physics, University of Science and Technology of China, Hefei, Anhui 230026, China}
\author{Yang Gao}
\email[Correspondence author:~~]{ygao87@ustc.edu.cn}
\affiliation{CAS Key Laboratory of Strongly-Coupled Quantum Matter Physics, and Department of Physics, University of Science and Technology of China, Hefei, Anhui 230026, China}
\affiliation{ICQD, Hefei National Laboratory for Physical Sciences at Microscale, University of Science and Technology of China, Hefei, Anhui 230026, China}
\affiliation{Hefei National Laboratory, University of Science and Technology of China, Hefei, Anhui 230026, China}
\affiliation{Anhui Center for Fundamental Sciences in Theoretical Physics, University of Science and Technology of China, Hefei 230026, China}
\author{Qian Niu}
\affiliation{CAS Key Laboratory of Strongly-Coupled Quantum Matter Physics, and Department of Physics, University of Science and Technology of China, Hefei, Anhui 230026, China}

\date{\today{}}

\begin{abstract}
We propose a magnon-driven anomalous Hall effect in altermagnets, arising from the coupling between coherently excited chiral magnons and chiral electronic motion. 
Using density-matrix perturbation theory and symmetry analysis, we show that the resulting Hall conductivity is solely determined by the chiralithy of the Néel-order precession, in sharp contrast to the anomalous Hall effect from the equilibrium Néel order. It then has distinct symmetry requirements from the latter and can exist even when the latter is forbidden by symmetry. 
The magnon-driven anomalous Hall effect is exemplified in a minimal lattice model with the same symmetry of the altermagnet CrSb, which hosts no static anomalous Hall effect. 
Our results reveal a direct interplay between chiral magnons and chiral electronic motion, paving the way of probing magnon chirality and to control electronic chirality through magnons.
\end{abstract}

\maketitle

In solid state physics, chirality is a fundamental concept that characerizes the circular motion of various particles, such as electrons\cite{Wan2011,Xu2015,Chang2018}, phonons\cite{Zhang2015,Zhu2018,Ishito2023,Ueda2023,Cheong2022,Chen2019}, and magnons\cite{Roessli2002,Nambu2020,Liu2022,ifmmodeSelseSfimejkal2023}. It can play important roles in a wide range of quantum phenomena, such as exciton formation\cite{Juan2017,Ma2017,Liu2019,He2020,Xu2020}, magnetic dynamics\cite{Zhang2019,Wang2024,Jin2025}, thermal Hall effect\cite{Katsura2010,Murakami2017,Tang2023}, and hence can stimulate technological developments in spintronics\cite{ifmmodecheckZelsevZfiutiifmmodeacutecelsecfi2004,Chumak2015,Yuan2022,Bozhko2020}, phononics\cite{Maldovan2013,Liu2020}, magneto-optical functionalities\cite{Buschow1983,2004,Haider2017}, and so on. In this regard, the coupling between different chiral modes is a central research theme relating to generation and detection of chirality. Specifically, the chiral phonon-magnon coupling and the chiral phonon-electron coupling have been studied intensively\cite{Zhang2019,Zhou2025,Weissenhofer2025,Xue2025,Wang2024}. For example, the hybrid magnon-phonon mode can show distinct chiral behavior\cite{Zhang2019}, and the valley exiton mode can be affected deeply by the chiral phonon mode\cite{Liu2019,He2020}. In comparison, a fundamental question remains largely untouched: can the chiral magnon mode couple to electrons and lead to distinct chiral transport behavior?

%
%
The chiral motion of electrons is most prominently manifested in the anomalous Hall effect\cite{Nagaosa2010}. Microscopically, the chirality is captured by the Berry curvature of the equilibrium ground state, which is necesarily associated with a macroscopic chiral direction in crystals, such as the ferromagnetic spin order\cite{Karplus1954,Birss1964,Yao2004}. Reversing the spin order then reverses the anomalous Hall response. Beyond equilibrium, the spin precession or the magnon mode offers another chiral direction, i.e., the precession direction. Therefore, it is an interesting and important question whether the anomalous Hall effect can be directly induced by such a dynamical chiral direction.

\begin{figure}
	\includegraphics[width=8.5cm,angle=0]{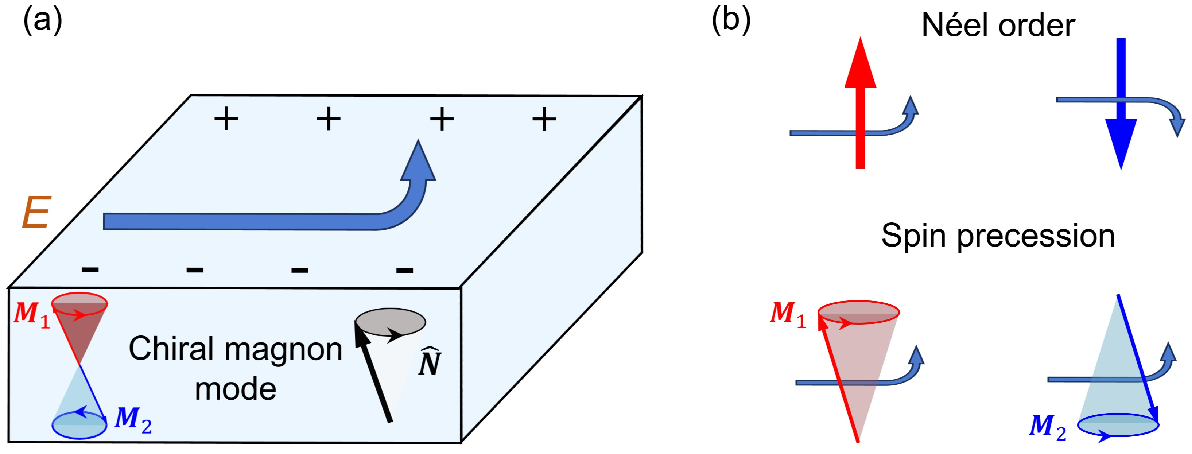}
	\caption{The magnon-driven anomalous Hall effect. (a) Hall-type response driven by Néel vector precession. (b) physical origin of the magnon-driven anomalous Hall effect: two chiral directions are present in altermagnets, the Néel order and its precession direction; the anomalous Hall effect can arise from either chiral direction but only in the later case, the opposite local spin orders contribute constructively.}
	\label{Fig1}
\end{figure}

In this work, we answer both questions in the affirmative, by showing that chiral magnon modes in altermagnets can generate a distinct anomalous Hall response~(as shown in Fig.~\ref{Fig1}(a)), in sharp contrast to that due to the equilibrium altermagnetic spin order. We chose altermagnets for two reasons: its low energy excitation contains magnon modes with both chiralities\cite{ifmmodeSelseSfimejkal2023,Jin2025,Liu2024,Alaei2025,Chen2025}, in sharp contrast to the magnon mode with single chirality in collinear ferromagnets\cite{Rezende2020}, and they are separated in energy, in sharp contrast to the degenerate magnon mode in $PT$- or $T\tau$-symmetric antiferromagnets\cite{Cheng2014,Li2020}. This magnon-driven anomalous Hall effect has a clear picture as shown in Fig.~\ref{Fig1}(b): at equilibrium, Berry curvatures associated with opposite local spin orders are also opposite, which contributes destructively to the anomalous Hall effect; however, opposite local spin orders precess in the same manner and hence share the same precession direction, which then contributes constructively to the anomlaous Hall effect.

Symmetry-wise, the magnon-driven anomalous Hall effect is also distinct from that due to the equlibrium spin order. It is linear in the preccession direction of the Néel vector as shown in Eq.~\eqref{eq:response2} but the corresponding coefficient is independent of the Néel vector. As a result, the flipping of the Néel order does not affect the response coefficient, consistent with the intuitive picture shown in Fig.~\ref{Fig1}(b). We offer systematic analysis to show such difference. Most strikingly, the magnon-driven anomalous Hall effect to exist even when the one due to the equilibrium spin order vanishes. 
As a concrete example, we consider a collinear antiferromagnetic model sharing the symmetry of altermagnetic CrSb\cite{Reimers2024,Ding2024}. Although being altermagnetic, its spin order does not support an anomalous Hall effect\cite{Urata2024,Ding2024}. However, its magnon mode can driven a Hall response, which shows a dipolar angular dependence with the rotation of spin order, in consistent with the spin-group analysis. Our results reveal the distinct interplay between chiral magnons and chiral electrons.

{\it Microscopic theory.---} To show the magnon-driven anomalous Hall effect, we first derive a steady current in response to both the electric field and spin precession, i.e.,
\begin{eqnarray}\label{eq:response1}
	j_i=\chi_{ijk\ell}(0,\omega_\alpha,\omega_\beta)E_j\hat{N}_k(\omega_\alpha)\hat{N}_\ell(\omega_\beta).
\end{eqnarray}
Here $\boldsymbol{E}$ is a static electric field, $\hat{\boldsymbol{N}}(t)=\hat{\boldsymbol{N}}(\omega)e^{-i\omega t} + \hat{\boldsymbol{N}}(-\omega)e^{i\omega t}$ which contains only the time-varying part of the Néel vector due to preccession, $\hat{\boldsymbol{N}}(-\omega)=\hat{\boldsymbol{N}}^*(\omega)$, and $\omega_{\alpha,\beta}=\pm \omega$. We note that the momentum of the magnon mode does not directly affects the current. It can, however, affects the chirality of the magnon mode, which is different for opposite momentums. 

We are interested in the Hall-type response. Therefore, the $ij$ indices shall be antisymmetrized. Moreover, the magnon mode is always circularly polarized. Therefore, the $k\ell$ indices should also be antisymmetrized and we shall take $\omega_\alpha+\omega_\beta=0$. With all these in mind, we can rewrite Eq.~\eqref{eq:response1} as follows:
\begin{eqnarray}\label{eq:response2}
	j_i=\varepsilon_{ija}\tilde{\chi}_{ab}(\omega)E_j \left[i\hat{\boldsymbol{N}}(\omega)\times\hat{\boldsymbol{N}}^*(\omega)\right]_b,
\end{eqnarray}
where the rank-four tensorial coefficient $\chi_{ijk\ell}$ essentially reduces to a rank-two tensorial coefficient
\begin{align}\label{eq:coefficient}
	\tilde{\chi}_{ab}(\omega)=\frac{i}{4}\varepsilon_{ija}\varepsilon_{k\ell b}\left[\chi_{ijk\ell}(0,-\omega,\omega)-\chi_{ijk\ell}(0,\omega,-\omega)\right].
\end{align}

The coefficient $\tilde{\chi}$ can be drived using the density-matrix perturbation theory. We start from the electric Hamiltonian in the presence of both a static electric field and the dynamical Néel-vector variation,
\begin{align}\label{eq:Ham_total}
	\hat{H}=\hat{H}_0+e\boldsymbol{E}\cdot\hat{\boldsymbol{r}}+\hat{F}_i\hat{N}_{i}(t),
\end{align}
where $\hat{H}_0$ is the unperturbed electronic Hamiltonian. The second and third terms represent perturbations from the electric field and Néel-order precession, respectively. Here, $\hat{F}_i\equiv{\partial \hat{H}_0}/{\partial \hat{N}_i}$ denotes the linear coupling to the Néel-vector deviation. Higher-order terms in $\hat{N}_i$ are ignored, as they are symmetric in $\hat{\bm N}$ and hence does not contribute to $\tilde{\chi}_{ab}$. Throughout this work, repeated indices imply summation.

The density matrix can then be obtained from the total Hamiltonian through the following kinetic equation:
\begin{align}\label{eq:eq_of_motion}
	i\hbar\frac{d \hat{\rho}}{d t}=[\hat{H},\hat{\rho}]-\frac{i\hbar}{\tau_0}\left(\hat{\rho}-\hat{\rho}^{(0)}\right),
\end{align}
where $\hat{\rho}^{(0)}$ is the equilibrium density matrix and $\tau_0$ denotes the phenomenological relaxation time of the electronic states. The last term implements relaxation within the standard relaxation-time approximation. We then solve Eq.~\eqref{eq:eq_of_motion} up to the third order to obtain $\delta\hat{\rho}^{(3)}$, in accordance with Eq.~{eq:response1}. The detail can be found in the supplementary materials. Morever, we note that the last term in Eq.~\ref{eq:Ham_total} arises from the exchange field, which commutes with $\hat{\boldsymbol{r}}$ and hence does not modify the velocity operator. The induced current then reads as
\begin{align}
	j_i=-e\text{Tr}\left(\delta\hat{\rho}^{(3)}\hat{v}_i\right),
\end{align}
with $\hat{v}_i=\partial \hat{H}_0/(\hbar\partial k_i)$ being the velocity operator. We can then obtain $\chi_{ijk\ell}$ and, through Eq.~\ref{eq:coefficient}, $\tilde{\chi}_{ab}$.

The magnon-driven anomalous Hall effect corresponds to the part of the response function that is time-reversal even. It is then even with respect to the equilibrium Néel vector $\hat{\bm N}_0$, i.e., $\tilde{\chi}_{ab}(-\hat{\boldsymbol{N}}_0)=\tilde{\chi}_{ab}(\hat{\boldsymbol{N}}_0)$. Its epxression can be found in the supplementary materials. The resulting current only relies on the chiral precession direction so that the opposite spin orders shall contribute constructively as shown in Fig.~\ref{Fig1}(b).

\begin{table}
	\caption{Symmetry constraints on the magnon-driven anomalous Hall conductivity tensor $\tilde{\chi}_{\rho r}$ under various symmetry operations. Here, $n, n' \ge 2$. The symbols $\checkmark$ and $\times$ denote symmetry-allowed and symmetry-forbidden components, respectively. $\tilde{C}_{n'\perp}$ indicates a rotation axis perpendicular to the $z$ axis. ``None" signifies the absence of a generalized rotational operation in the parent point group of the magnetic point group.}
	\begin{ruledtabular}
		\begin{tabular}{cccc}
			& $\tilde{C}_{nz}$                         & $\tilde{C}_{nz}$ and $\tilde{C}_{n'\perp}$     & None         \\ \hline
			$\tilde{\chi}_{xx}$, $\tilde{\chi}_{yy}$, $\tilde{\chi}_{zz}$  &$\checkmark$                      & $\checkmark$                   & $\checkmark$ \\  
			$\tilde{\chi}_{xy}$, $\tilde{\chi}_{yx}$                       &$\checkmark$                      & $\times$                       & $\checkmark$ \\
			$\tilde{\chi}_{xz}$, $\tilde{\chi}_{yz}$, $\tilde{\chi}_{zx}$, $\tilde{\chi}_{zy}$  & $\times$    & $\times$                       & $\checkmark$          
		\end{tabular}
	\end{ruledtabular}
	\label{table-1}
\end{table}

{\it Symmetry analysis.---} The most striking feature of the magnon-driven anomalous Hall effect is that it has quite different symmetry requirements compared with the anomalous Hall effect from the equilibruim spin order, since the time reversal operation does not affect the former\cite{Onsager1931,Onsager1931a,McClarty2024,Xiao2025}. Specifically, the response coefficient $\tilde{\chi}_{ab}$ is subjected to a derived group obtained from the magnetic point group by further identifying the time reversal operation with the identity. This derived group does not, in general, coincide with the crystallographic point group. Moreover, the inversion operation also leaves $\tilde{\chi}_{ab}$ unchanged. Therefore, for $\tilde{\chi}_{ab}$, $C_n$ and $IC_n$ are equivalent and we will unify them by $\tilde{C}_n$. In this notation, $\tilde{C}_2$ contains the mirror operation.  Without loss of generality, we choose the $z$ axis as the principal rotation axis. The symmetry-allowed components of $\tilde{\chi}_{ab}$ are summarized in Table~\ref{table-1}.

Notably, the diagonal components $\tilde{\chi}_{xx}$, $\tilde{\chi}_{yy}$, and $\tilde{\chi}_{zz}$ are allowed under arbitrary rotational symmetry. In other words, with the spin precession from the magnon mode in altermagnets, there should always a Hall-type response in the plane perpendicular to the precession direction. This demonstrates the robustness of the magnon-driven anomalous Hall effect.
If only a single rotation axis $\tilde{C}_{nz}$ with $n\ge 2$ is present, additional off-diagonal components such as $\tilde{\chi}_{xy}$ and $\tilde{\chi}_{yx}$ are also allowed. 

Depending on whether the spin order lies perpendicular to or parallel with the Hall deflection plane, the anomlaous Hall effect has two versions: the out-of-plane anomalous Hall effect as consistent with the empirical law\cite{Pugh1953,Kundt1893,Pugh1930,Pugh1932} and the in-plane anomalous Hall effect when the crystalline symmetry is low or the Hall deflection plane has a high index\cite{Tan2021,Cao2023,Zhou2022,Liu2025}. The anomalous Hall effect induced by magnon and that by spin order have the same symmetry requirements for the later~(i.e., no $C_2$ axis along the spin order and perpendicular to the Hall deflection plane) but they are different in symmetry for the former. In altermagnets, if the magnetic point group of a crystals contains $C_nT$ with $n\ge 2$, the Hall deflection plane in the anomalous Hall effect from the spin order cannot be perpendicular to the rotation axis, which then forbids the out-of-plane configuration. However, such configuration is still allowed for the magnon-driven anomalous Hall effect, in which case, the spin preccession direction, the spin order direction, and the Hall conducitvity vector are all along the rotation axis.

For example, in $\mathrm{FeSe}_2$, a $d$-wave altermagnet with Néel vector along $y$ and symmetry $TC_{2y}$, the Hall conductivity vector $\bm \sigma^H$ cannot be along $y$-direction, whereas a finite $\sigma^H_y$ can be induced when the spin order precesses about $y$ axis\cite{Liu2025,Attias2024}. Similarly, in $\mathrm{CrSb}$, a $g$-wave altermagnet with Néel vector along $z$ and symmetry $TC_{6z}$, $\bm \sigma^H$ from spin order vanishes identically\cite{Urata2024,Ding2024}. However, it can be induced by magnons.

With the help of the spin group\cite{ifmmodeSelseSfimejkal2022,Liu2022a,Chen2024,Xiao2024,Jiang2024,Liu2025,Liu2026}, we can also identify the structure of the magnon-driven anomalous Hall effect in the Néel-vector space. For this purpose, we differentiate the spin frame from the lattice frame. The latter is determined by the crystal axis while the former rotates with the spin order. We further assume that the $z$-axis in the spin frame is always parallel with the Néel order. We then treat the response coefficient with both the spin frame and the lattice frame, and re-label this version by $\tilde{\chi}_{ab}^{\rm sep}$ to differentiate it from $\tilde{\chi}_{ab}$  in the lab frame as defined in Eq.~\eqref{eq:response2}. For $\tilde{\chi}_{ab}^{\rm sep}$, the first index is related to the Hall-type deflection and should then transform with the lattice frame, while the second index is related to the spin precession and should then transform with the spin frame.  Since the spin order always precesses about its equilibrium direction for low-energy excitations, we are only interested in the $\tilde{\chi}_{az}^{\rm sep}$ component.  

The anisotropy of $\tilde{\chi}_{az}^{\rm sep}$ in the Néel-vector space  originates from the spin-group symmetry breaking. To proceed, we treat the spin–orbit coupling perturbatively and expand $\tilde{\chi}_{az}^{\rm sep}$ with respect to the spin-orbit vector $\ell_j^m$ which characterizes the relative orientation between the spin frame and the lattice frame. The expansion coefficient is then subject to the constraint of spin group. For altermagnets, the spin-only symmetry is generated by $TC_{2x}^{s}$ and $C_{\infty z}^{s}$, where $z$ and $x$ are in the spin frame. Since $C_{2x}^s$ flips the $z$ direction in the spin frame, the response coefficient transforms as
\begin{align}
	\tilde{\chi}_{az}^{\rm sep} \xrightarrow{TC_{2x}^s} -\tilde{\chi}_{az}^{\rm sep}\,, \tilde{\chi}_{az}^{\rm sep}  \xrightarrow{C_{\infty z}^s} \tilde{\chi}_{az}^{\rm sep} \,.
\end{align}
The first constraint necessarily forbids the zeroth-order term in the expansion of $\tilde{\chi}_{az}^{\rm sep}$.  At first order, $\tilde{\chi}_{az}$ should only depend on components of $\ell_j^m$ with $m=z$ due to the second constraint, yielding 
\begin{align}\label{eq_exp}
	\tilde{\chi}_{iz}^{\rm sep}=\alpha_{ij}\ell^z_j=\alpha_{ij} \hat{N}_j\,,
\end{align}
where $\alpha_{ij}$ is at first order of the strength of the spin-orbit coupling. The last equality holds since the $z$ direction in the spin frame is parallel with the Néel vector.
This immediately shows the role of the spin-orbit coupling in the magnon-driven anomalous Hall effect.

The coefficient $\alpha_{ij}$ is further constrained by the nontrivial spin group. In collinear antiferromagnets, the nontrivial spin group element takes the form $\{C^{s}_{2\perp}|G^{L}\}$ and $\{E^{s}|G^{L}\}$, where $C^{s}_{2\perp}$ denotes a twofold spin rotation about an axis within the $xy$ plane in the spin frame, $E^{s}$ is the identity in spin space, and $G^{L}$ represents a  crystalline point-group operation. Because $C^{s}_{2\perp}$ simultaneously flips the sign of $\tilde{\chi}_{az}^{\rm sep}$ and $\ell^{z}_{j}$, $\alpha_{ij}$ is only affected by $G^L$. 

It is instructive to compare the expansion in Eq.~\eqref{eq_exp} with that of the anomalous Hall conductivity in altermagnets. The latter has a similar form\cite{Liu2025}:
\begin{align}
	\sigma_i^H=\beta_{ij} \ell_j^z\,.
\end{align}
$\beta_{ij}$ and $\alpha_{ij}$ transform in the same way under the spin-only group. However, for the nontrivial spin group, $C_{2\perp}^s$ affects $\beta_{Ij}$ but not $\alpha_{ij}$. This is the symmetry reason for the robustness of the magnon-driven anomalous Hall effect.

Finally, we comment that the magnon-driven anomalous Hall effect can also be generated to $\mathcal{PT}$-symmetric or $\mathcal{T}\tau$-symmetric collinear antiferromagnets. In both cases, the antiferromagnetic spin order cannot support the anomalous Hall effect. However, if magnons with fixed chirality can be excited through, e.g., the spin current injection, a nonzero anomalous Hall effect can emerge from $\tilde{\chi}_{ab}$.

\begin{figure}
	\includegraphics[width=8.5cm,angle=0]{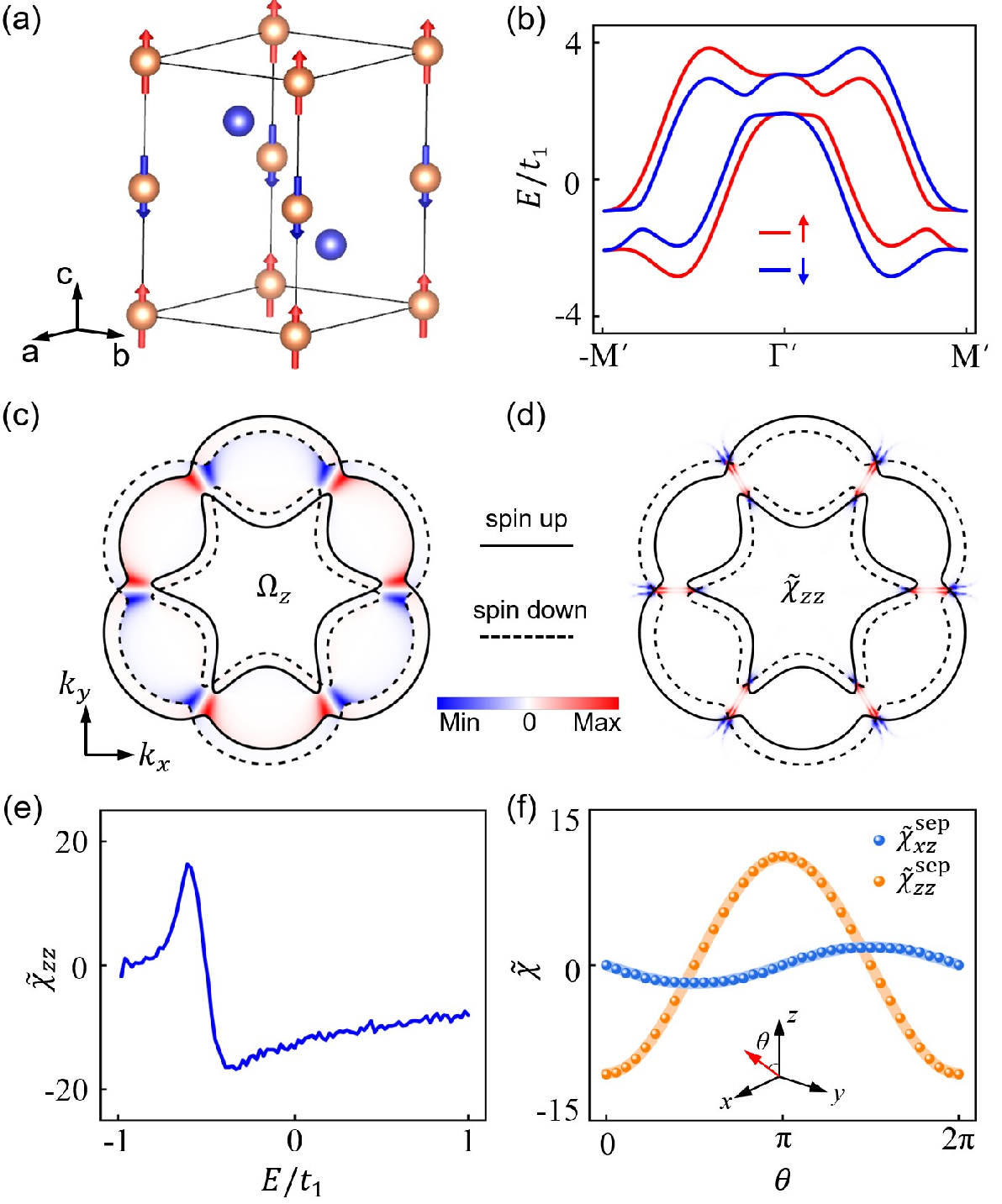}
	\caption{Magnon-driven anomalous Hall effect in an altermagnetic minimal model. (a) The lattice structure. (b) Band structure of the minimal model in the absence of spin-orbit coupling. The $k$-path are $-M'=(-0.5,0,0.25)$, $\Gamma'=(0,0,0.25)$, and $M'=(0.5,0,0.25)$.  (c), (d) Spin-up and spin-down Fermi surfaces in $k_x-k_y$ plane at $k_z=\pi/2$ (without spin-orbit coupling), together with the Berry curvature $\Omega_z(\boldsymbol{k})$ in (c) and the magnon-driven anomalous Hall response coefficient $\tilde{\chi}_{zz}(\boldsymbol{k})$ in (d) (arbitrary units). (e) Magnon-driven anomalous Hall coefficient $\tilde{\chi}_{zz}$ as a function of Fermi energy with the Néel vector along the $z$ direction. (f) Angular dependence of $\tilde{\chi}_{xz}^{\rm sep}$ and $\tilde{\chi}_{zz}^{\rm sep}$ as the Néel vector rotates from the $z$ to $x$ axis, parameterized by the angle $\theta$. The response coefficient is measured in units of $c_0$S/cm, with $c_0=\tau_0/\hbar$. The inset shows the Néel-vector direction. Model parameters are $t_2/t_1=1.0$, $t_3/t_1=0.5$, $\mu/t_1=0.5$, $\lambda/t_1=0.004$, and $J/t_1=0.45$. The magnon frequency is $\omega/t_1=0.1$. The Fermi energy is fixed at $E_F/t_1=0$ in (f).}
	\label{Fig2}
\end{figure}

{\it Lattice Model.---} To illustrate the magnon-driven anomalous Hall effect, we consider a minimal lattice model capturing the essential symmetry properties of the altermagnet CrSb (structure in Fig.~\ref{Fig2}(a)). The Hamiltonian reads\cite{Roig2024,GonzalezHernandez2025}
\begin{align}\label{eq:Ham}
	H=\epsilon_{0,k}+t_{x,k}\tau_x+t_{z,k}\tau_z+\tau_y\boldsymbol{\lambda}\cdot\boldsymbol{\sigma}+\tau_z\boldsymbol{J}\cdot\boldsymbol{\sigma},
\end{align}
where the Pauli matrices $\boldsymbol{\tau}$ act on sublattices and $\boldsymbol{\sigma}$ on spin. The momentum-dependent coefficients are given by $\epsilon_{0,k}=t_1(\cos k_x+2\cos k_x/2\cos \sqrt{3}k_y/2)+t_2 \cos k_z-\mu$, $t_{x,k}=t_2\cos k_z/2$, $t_{z,k}=t_3\sin k_z f_y(f_y^2-3f_x^2)$, $\lambda_{x,k}=\lambda \cos k_z/2(f_x^2-f_y^2)$, $\lambda_{y,k}=-2\lambda\cos k_z/2f_xf_y$, $\lambda_{z,k}=\lambda \sin k_z/2f_x(f_x^2-3f_y^2)$, $f_x=\sin k_x+\sin k_x/2\cos \sqrt{3}k_y/2$, and $f_y=\sqrt{3}\cos k_x/2\sin \sqrt{3}k_y/2$. We have set the lattice constant to be $1$. The crystallographic point group for this model is $D_{6h}$ consistent with the hexagonal lattice structure. The spin point group is generated by $\{C_{2\perp}^{s}|C_{6z}^{L}\}$, $\{C_{2\perp}^{s}|\sigma_{xy}^{L}\}$, and $\{E^{s}|C_{2x}^{L}\}$, consistent with the spin structure of CrSb\cite{Reimers2024,Ding2024}. 

In Fig.~\ref{Fig2}(b), we plot the band structure in the absence of spin–orbit coupling. Although the magnetic moments are collinear and antiparallel, fully compensating each other, the bands are clearly spin-split, reflecting the characteristic nature of altermagnetism. Figures~\ref{Fig2}(c) and \ref{Fig2}(d) show the spin-up (solid) and spin-down (dashed) Fermi surfaces in a Brillouin-zone slice. They are generally separated in the momentum space and are consistent with the $g$-wave nature of ${\rm CrSb}$\cite{ifmmodeSelseSfimejkal2022}. In Fig.~\ref{Fig2}(c), the Berry curvature distribution is also shown by color. Its sign alternates in the azimuthal direction since the positive and negative value are associated with the spin-up and spin-down bands, respectively, leading to an overall cancellation. By contrast, the $k$-resolved response coefficient $\tilde{\chi}_{33}$ exhibits identical distributions for the spin-up and spin-down bands and does not change sign in the azimuthal direction~(Fig.~\ref{Fig2}(d)). This behavior demonstrates that the magnon-driven anomalous Hall response is governed by the magnon chirality rather than by the local spin order, and confirms the physical picture in Fig.~\ref{Fig1}(b).

As discussed previously, the magnetic point group of CrSb forbis the anomalous Hall effect\cite{Urata2024,Ding2024}, which is also true for our lattice model when the Néel order is along the $z$ direction. However, $\tilde{\chi}_{zz}$ is nonzero as shown in Fig.~\ref{Fig2}(e), indicating the existence of the magnon-driven anomalous Hall effect. It also depends sensitively on the Fermi energy and can switch sign across the band crossing point.

In Fig.~\ref{Fig2}(f), we show the dependence of the magnon-driven anomalous Hall effect on the direction of the equilibrium Néel vector. In CrSb, when the Néel vector deviates from the $z$ axis, the magnetic point group can be greatly supressed, which allows a nonzero anomalous Hall effect associated with the Néel order. However, a detailed analysis shows that the corresponding Hall conductivity vector shall start at the cubic order of the Néel vector~(and equivalently the strength of the spin-orbit coupling). In contrast, the magnon-driven anomalous Hall effect can start at the linear order. Specifically, for the three generators of the nontrivial spin group, only the orbit part, i.e., $C_{6z}^L$, $\sigma_{xy}^L$, and $C_{2x}^L$ affect the coefficient $\alpha_{ij}$ in Eq.~\eqref{eq_exp}. Therefore, the nonzero elements are $\alpha_{xx}=\alpha_{yy}$ and $\alpha_{zz}$. Rotating the Néel vector from the $z$ axis toward the $x$ axis by an angle $\theta$ yields
\begin{align}\label{eq:CrSb}
	\tilde{\chi}_{xz}^{\rm sep}=\alpha_{xx}\sin\theta, \notag\\
	\tilde{\chi}_{zz}^{\rm sep}=\alpha_{zz}\cos\theta.
\end{align}
These are in good agreement with the curve in Fig.~\ref{Fig2}(f), confirming the previous symmetry analysis.

{\it Summary.---} In this work, we identify and characterize a novel transport phenomenon—the magnon-driven anomalous Hall effect. The resulting Hall conductivity vector only depends on the chirality of spin-order precession instead of the Néel order. Therefore, it has distinct symmetry requirements and can be nonzero even when the Néel order itselt cannot support an anomalous Hall effect. It generally has a dipolar structure in the Néel vector space. Our results reveals the close coupling between chiral magnon modes and chiral electronic motion and provide a route to converting magnonic signals into electronic responses and enabling the detection of magnon chirality via transport measurements, which provides opportunities for exploiting chiral magnons in future spintronic applications.

\begin{acknowledgments}
The authors are supported by the National Key R${\rm \&}$D Program under grant Nos. 2022YFA1403502 and the National Natural Science Foundation of China (12234017). Y. G. is also supported by the National Natural Science Foundation of China (12374164). Z. L. is also supported by Postdoctoral Fellowship Program of CPSF (GZC20232562), fellowship from the China Postdoctoral Science Foundation (2024M753080), and National Natural Science Foundation of China (12504066). Y. G. and Q. N. are also supported by the Innovation Program for Quantum Science and Technology (2021ZD0302802). The supercomputing service of USTC is gratefully acknowledged.
\end{acknowledgments}

\bibliography{MAHE}

\end{document}